\pdfoutput=1
\def\year{2018}\relax
\documentclass[letterpaper]{article} %DO NOT CHANGE THIS
\usepackage{aaai18}  %Required
\usepackage{times}  %Required
\usepackage{helvet}  %Required
\usepackage{courier}  %Required
\usepackage{url}  %Required
\usepackage{graphicx}  %Required
\usepackage{graphicx}
\graphicspath{ {images/}}
\begin{document}
% The file aaai.sty is the style file for AAAI Press 
% proceedings, working notes, and technical reports.
%

%\nocite{*}
%%%%% Commands
\newcommand{\one}{({\em i}\/)}
\newcommand{\two}{({\em ii}\/)}
\newcommand{\three}{({\em
iii}\/)}
\newcommand{\four}{({\em iv}\/)}
\newcommand{\five}{({\em v}\/)}
\newcommand{\six}{({\em vi}\/)}
\newcommand{\seven}{({\em vi}\/)}

\def\tofill{\textcolor{red}{XX}\xspace}
\def\tofillp{\textcolor{red}{XX\%}\xspace}

\def\eg{\emph{e.g. }}
\def\etc{\emph{etc. }}
\def\ie{\emph{i.e.,}}
\def\etal{\emph{et al. }}
\def\vs{\emph{vs.}\xspace}
\def\cf{\emph{cf.}\xspace}
%\definecolor{wscolor}{RGB}{26, 26, 255}
%\newcommand\ws[1]{  \textcolor{wscolor}{#1}} %for websites

%\makeatletter
%\newcommand{\myconst}[1]{#1\renewcommand{\@currentlabel}{#1}}

%%%%%%%% Data by Numbers Dataset %%%%%%%%%
\def\NumberOfStreamingLinksInDataset{795,698}
\def\NumberOfStreamingLinkCrawalwdFurther{471,495}
\def\NumberOfStreamingLinkCrawalwdFurtherPercent{59.3\%}
\def\TotalNumberOfVideos{139,335}
\def\TotalNumberOfVideosWithStreamingLinks{129,809}
\def\AverageNumberOfStreamingLinkPerVideo{6}
\def\NumberOfCyberlockers{151}

\def\NumberOfAutonomousSystem{9}
\def\NumberOfCountry{8}
\def\NumberOfContinent{2}
\def\NumberOfStreamingDomain{33}
\def\NumberOfStreamingServers{1,903}

%%%%%Lumen Data by Numbers%%%%%%%%%%%%%
\def\TotalLumenNotice{1,341,365}
\def\TotalInfringingURL{781,763,167}

\def\TotalIdentifiedInfringingURL{21,801,381}
\def\NumberOfNoticeFound{49,829}
\def\NumberReportingOrganisationFound{304}

\def\TotalIdentifiedInfringingURLPercent{2.8\%}
\def\NumberOfStreamingLinkIdentifiedInLUmen{2,669}
\def\NumberOfVideosIdentified{275} %2017videos

\title{Movie Pirates of the Caribbean: Exploring Illegal Streaming Cyberlockers}

\author{Damilola Ibosiola\textsuperscript{\textdagger}, Benjamin Steer\textsuperscript{\textdagger}, Alvaro Garcia-Recuero\textsuperscript{\textdagger}, Gianluca Stringhini\textsuperscript{\textdaggerdbl},\\{\Large\bf Steve Uhlig\textsuperscript{\textdagger} \and Gareth Tyson\textsuperscript{\textdagger}}\\
\textsuperscript{\textdagger}Queen Mary University of London, \textsuperscript{\textdaggerdbl} University College London\\
\textsuperscript{\textdagger}\{d.i.ibosiola, b.a.steer, alvaro.garcia-recuero, steve.uhlig, g.tyson\}@qmul.ac.uk, \textsuperscript{\textdaggerdbl}g.stringhini@ucl.ac.uk
}
% \author{Anonymous Author(s)}

\maketitle

\begin{abstract}
Online video piracy (OVP) is a contentious topic, with strong proponents on both sides of the argument. Recently, a number of illegal websites, called \emph{streaming cyberlockers}, have begun to dominate  OVP. These websites specialise in distributing pirated content, underpinned by third party indexing services offering easy-to-access directories of content. This paper performs the first exploration of this new ecosystem. It characterises the content, as well the streaming cyberlockers' individual attributes. We find a remarkably centralised system with just a few networks, countries and cyberlockers underpinning most provisioning. We also investigate the actions of copyright enforcers. We find they tend to target small subsets of the ecosystem, although they appear quite successful. 84\% of copyright notices see content removed.  
\end{abstract}

%%%%%%%%%%%%%%%%%%%%%%%%%%%%%%%%%%%%%%%%%%%%%%%%%%%%%%%%%%%%%%%%%
\section{Introduction}
%%%%%%%%%%%%%%%%%%%%%%%%%%%%%%%%%%%%%%%%%%%%%%%%%%%%%%%%%%%%%%%%%

%%%%%%%%%%%%%%%%%%% OVERALL VIEW %%%%%%%%%%%%%%%%%%%
%Online Video Piracy (OVP) is a contentious topic, with strong proponents on both sides of the argument. 
Online Video Piracy (OVP) has been the focus of an increasing debate over the past years.
Entire political movements have emerged around the idea that content should be freely available~\cite{li2009pirate}, whilst lobbyists consistently argue that dire consequences exist. For example, CBP reported that piracy costs the US economy over 750,000 jobs, and between \$200-250B per year~\cite{kalchris2012}. Regardless of one's stance, it is undeniable that OVP constitutes a major web traffic generator~\cite{MarkMonitor2011,robert2016}, and creates significant interest from users, law enforcers and the creative industries alike.
%According to \emph{Computer.org}~\cite{ortiz2011peer}, the use of P2P declined by 71\% between the years 2007-2009 alone.

% \begin{figure}[t] \includegraphics[width=\linewidth]{trend} \centering
% \caption{Worldwide comparison of\emph{bittorrent}, \emph{openload} and
% \emph{putlocker} on google trends from Jan-2004 to Oct-2017} \label{fig:trend}
% \end{figure}

Traditionally, online piracy was dominated by decentralised peer-to-peer (P2P) systems such as Gnutella and BitTorrent. However, these have since been surpassed by a new breed of more centralised service allowing users to stream pirated content directly from YouTube-like websites --- so called \emph{streaming cyberlockers}. These streaming cyberlockers have gained huge traction. For example, many prominent portals are in the Alexa Top 1K, \eg \emph{openload.co}, \emph{thevideo.me} and \emph{vidzi.tv}. Their ease of use attracts a large number of users and the difficulties law enforcers encounter when detecting user identities provides viewers with relative safety from prosecution. Organisations such as the Motion Picture Association of America (MPAA) have, therefore, shifted their efforts towards shutting down the cyberlockers themselves. Examples of prominent shutdowns witnessed in this paper include \emph{allmyvideos.net}, \emph{vidbull.com} and \emph{vodlocker.com}.

Although similar to typical social video platforms, these streaming cyberlockers address a very different need. They employ few, if any, copyright checks and utilise evasion tactics to avoid detection. For example,
they often curate content on their front-pages to appear legitimate and disable search to prevent visitors from looking up videos. This has created an interesting ecosystem where cyberlockers depend on third party (crowd-sourced) \emph{indexing websites} that create a searchable directory of direct links (URLs) to the videos. These two types of website operate hand-in-hand with a symbiotic relationship, collectively underpinning a global network of online piracy.

To date, little is known about this emerging ecosystem. Its exploration, however, could reveal a range of insights regarding how large-scale copyright infringement takes place. This raises several particularly interesting questions, including: what type of copyright content is shared?  What are the dynamics regarding both content and website appearance/disappearance? What web hosting characteristics are commonly seen and how resilient are they? How are these websites pursued by copyright enforcers and how do the websites react? 
%This encompasses a number of social factors, related to how users upload and access this content across sites.

To answer these questions we exploit several measurement methodologies
(\S\ref{sec:dataset}), acquiring evidence of the characteristics
exhibited in this domain. As it would be impossible to inspect the entire copyright infringement ecosystem, we have taken a slice of 3 prominent indexing sites, as well as \NumberOfStreamingDomain~different cyberlockers.
Between January and September 2017 we performed monthly crawls, collecting
all published videos on these indexing sites. In parallel, we have scraped their related cyberlockers, collecting data on each video, including its availability and where it is hosted. To complement this data we further gathered metadata on the videos themselves, \eg release date and genre. Finally, we have monitored legal take down notices, allowing us to understand the reaction of the cyberlockers to complaints.

%%%%%%%%%%%%%%%%%%% CONTRIBUTIONS %%%%%%%%%%%%%%%%%%%
We begin our analysis by exploring the streaming links shared on indexing sites (\S\ref{sec:char_links}). We find a set of web platforms actively involved in aggressive copyright infringement. Predominantly content is made up of recently released Drama, Comedy, Thriller, and Action films. However, we also observe a non-negligible amount of older content --- some videos are from over 100 years ago. The websites we monitor show clear temporal trends with periods of activity, followed by collapse --- likely driven by legal take downs. For example, \emph{putlocker.is} (an indexing site) ceased uploading new links three months into our measurements. This reveals a model rather more vulnerable than the decentralised P2P networks.

We then inspect the characteristics of the individual cyberlockers (\S\ref{sec:relationships}). We model these concepts as several graphs that capture the related attributes of websites. A key finding is the apparent centralisation of these portals, with a small set of dependencies vulnerable to attack from copyright enforcers. For example, we observe that 58\% of all videos are located within just two hosting providers (despite being spread across 15 cyberlockers). Similarly, we find strong signs that individual pirates tend to operate \emph{multiple} websites. For instance, although seemingly different cyberlockers, \emph{daclips}, \emph{gorillavid} and \emph{movpod} are all operated by the same owner. These three cyberlockers alone host 15\% of observed content. Again, this suggests a distribution model that is far less resilient than its decentralised P2P counterparts.

%For example, we find that these websites have a far higher number of monetisation plugins than seen in other domains, including both ad networks and cyrpto-currency miners. They also have distinct hosting patterns, with a preference for countries with lax copyright enforcement regulation. 

Finally, we inspect the behaviour of copyright enforcers (\S\ref{sec:removals}). By studying the takedown notices placed against the cyberlockers under observation, we find that most enforcers take a bulk approach --- selecting a set of cyberlockers and generating many notices. That said, most cyberlockers \emph{do} appear to placate such enforcers. During our measurement period, 84\% of notices later saw the content removed. Our results have implications for understanding modern copyright infringement both from the perspectives of content pirates and law enforcers (\S\ref{sec:conclusion}).

%%%%%%%%%%%%%%%%%%%%%%%%%%%%%%%%%%%%%%%%%%%%%%%%%%%%%%%%%%%%%%%%%
\section{Background \& Related Work} \label{sec:background}
%%%%%%%%%%%%%%%%%%%%%%%%%%%%%%%%%%%%%%%%%%%%%%%%%%%%%%%%%%%%%%%%%

Before beginning our analysis, we provide a brief overview of
the the general area, as well as related works.

\subsection{Overview of Video Piracy Stakeholders}

%An overview of the distribution process for pirated videos is shown in Figure~\ref{fig:ecosystem}. 
There are three major stakeholders worth considering. The failure of any of them would result in the collapse of the ecosystem. The players are:

% \begin{figure}[ht!]
% \centering 
% \includegraphics[trim= 2cm 17cm 0cm 0cm, width= 9.5cm, height=1.5cm]{ecosystem}
% \caption{Online pirated video distribution process through streaming cyberlockers. \gareth{Let's sit down and remake this figure, it's messy and difficult to parse. }}
% \label{fig:ecosystem} 
% \end{figure}

\vspace{2pt} \noindent\emph{Video Uploader:} A video uploader harvests video content (\eg using BitTorrent) and uploads it to a streaming cyberlocker. For each video uploaded, a unique URL is received. These URLs are published by the uploader on an indexing site with the appropriate metadata for searching.

\vspace{2pt} \noindent\emph{Streaming Cyberlocker:} A streaming cyberlocker is a web platform where a video uploader stores content. Typically a streaming site
is neither searchable nor indexed by search engines.
Users require the specific URL to view the content.
% To ensure a complete streaming process, streaming sites rely on third party video player softwares (JWPlayer and VideoJS) to embed and play videos.

\vspace{2pt} \noindent\emph{Indexing Site:} Indexing sites operate as a public directory, mapping video metadata (\eg title) to a list of cyberlocker URLs
where the content can be viewed. They allow viewers to search for any desired video and select a preferred streaming site.

\subsection{Related Work}

Online video distribution is not a new topic. 
The streaming cyberlockers work on a model of third parties uploading content. There are a range of video platforms allowing users to upload and share their own content, \eg YouTube~\cite{ZinkKyoungwon2008,TorresFinamore2011,MeeyoungKwak2007} and Vimeo~\cite{sastry2012tell}. Ding \etal characterised YouTube
uploader behaviour and classified the uploads~\cite{YingkaiDING2011}. It was discovered that the majority of content was copied and little actually user generated. Of most relevance to our work is the use of such platforms to distribute copyrighted material. There have been several studies looking at how platforms have been exploited for such purposes~\cite{clay2011blocking,hilderbrand2007youtube}. In response, platforms like YouTube now employ signature-based detection to prevent copyrighted material remaining online~\cite{dutta2008detecting}. This has led to a range of unusual evasion techniques, \eg removing portions of the film and injecting artefacts.

This complexity has resulted in pirated content moving away from these portals towards what are known as cyberlockers or one click file hosts (OCFH). These services offer remote storage, allowing users to share files. \cite{Arlitt2012} provide an understanding of the nature of OCFHs and their effect on the network. Sanju\`as-Cuxar \etal~ also analysed HTTP traffic emanating from OCFHs, ranking them amongst the major  contributors of HTTP traffic on the Internet~\cite{Sanju2013}.  Perhaps closest to our own work is \cite{Szydlowski2013,Chaabane2015,farahbakhsh2013investigating}. The first works scraped data from several OCFHs, such as MegaUpload and RapidShare, to understand the fraction of files that infringe copyright, whilst the second work investigated the impact of the MegaUpload shutdown on BitTorrent. Although closely related, our focus is not on \emph{file sharing} but on pirated \emph{video streaming}. We know of only one work targeting streaming services~\cite{Goethem2016}. This work investigated the security implications of illegal sports streaming, as well as how deceptive adverts and malware are used for monetisation. These sports sites are quite different to the movie sites we observe, primarily because they are \emph{live} broadcasts. Hence, we proceed to study the broader aspects of video piracy. Our paper sheds light on the behaviour of these websites in reaction to legal action, as well as the individual characteristics and relationships between them. To the best of our knowledge, this is the first paper focusing on the streaming cyberlocker ecosystem.

% \gareth{These following things appear to be reports and media articles? Don't
% just list these things - you need to talk about them and explicitly state how
% they relate to your own work. And importantly how they do not fulfil your
% goals.} \cite{ref14}, \cite{ref15} investigated the revenue source and operating
% cost of cyberlockers to project their income as they incur no cost for
% the content they distribute. Other related studies looked at their activities
% from a legal and protected content distribution angle; \cite{ref16},
% \cite{ref17} traces the initial footprint of TV piracy and efforts made to
% defeat the trend, they also highlighted that the presence of proper legislation
% has not curb the menace.

% \begin{table*}[ht!] \centering \begin{tabular}{ l|l|l|l|l } \hline Indexing Site
% & \begin{tabular}[c]{@{}l@{}}No. of \\ indexed videos\end{tabular} &
% \begin{tabular}[c]{@{}l@{}}No. of videos \\ with streaming links\end{tabular} &
% \begin{tabular}[c]{@{}l@{}}No. of \\ streaming links\end{tabular} &
% \begin{tabular}[c]{@{}l@{}}\% of videos \\ with streaming links\end{tabular} \\
% \hline putlocker.is & 25,700 & 24,974 & 148,878 & 97.2 \\ \hline watchseries.gs
% & 49,614 & 49,522 & 300,296 & 99.8 \\ \hline vodly.cr & 64,021 & 55,313 &
% 346,524 & 86.4 \\ \hline\hline Total & 139,335 & 129,809 & 795,698 & 93.2 \\
% \hline

% \end{tabular} \caption{Snapshot of indexed videos and stream links \gareth{Add number of unique cyberlockers seen and average/median number of links per page}}
% \label{table:1} \end{table*}

\begin{table*}[ht!]
\centering
\begin{tabular}{@{}llllll@{}}
\hline
Indexing Site  & \begin{tabular}[c]{@{}l@{}}No. of \\ indexed  videos\end{tabular} & \begin{tabular}[c]{@{}l@{}}No. of videos\\ with streaming links\end{tabular} & \begin{tabular}[c]{@{}l@{}}No. of\\ streaming links\end{tabular} & \begin{tabular}[c]{@{}l@{}}\% of videos\\ with streaming links\end{tabular}  & \begin{tabular}[c]{@{}l@{}}No. of unique \\ cyberlockers\end{tabular} \\ \hline
putlocker.is   & 25,700                                                            & 24,974                                                                       & 148,878                                                          & 97.2                                                                                                                                                  & 104                                                                    \\
watchseries.gs & 49,614                                                            & 49,522                                                                       & 300,296                                                          & 99.8                                                                                                                                                  & 125                                                                   \\
vodly.cr       & 64,021                                                            & 55,313                                                                       & 346,524                                                          & 86.4                                                                                                                                                  & 84                                                                    \\ \hline
Total          & 139,335                                                           & 129,809                                                                      & 795,698                                                          & 93.2                                                                                                                                               & 151                                                                   \\ \hline
\end{tabular}
\caption{Summary of data collected from each indexing site. }
\label{table:1}
\end{table*}

% \gareth{Gianluca made a good point - for the NAs, we can replace them with the average computed over the total, and the total number of unique cyberlockers} 

%%%%%%%%%%%%%%%%%%%%%%%%%%%%%%%%%%%%%%%%%%%%%%%%%%%%%%%%%%%%%%%%
\section{Methodology \& Data Collection}\label{sec:dataset}
%%%%%%%%%%%%%%%%%%%%%%%%%%%%%%%%%%%%%%%%%%%%%%%%%%%%%%%%%%%%%%%

We begin by presenting our measurement methodology. Our measurements follow three steps: \one~Collecting all streaming links from the indexing sites; \two~Visiting the links to check the availability of the videos; and \three~Gathering extended metadata for each video and website under study.

\subsection{Indexing Sites}

Due to the sheer number of indexing websites, it is impossible to evaluate them all. Hence the first step is to select a subset of indexing sites --- these operate as ``seeds'' which allow us to identify key cyberlockers. To achieve this, we inspected court orders obtained by the MPAA to understand those sites viewed as important by copyright enforcers. We then complemented this by performing a variety of searches on Google using relevant terms (\eg ``free films'', ``watch movies free''). This was intended to discover websites that a typical user may encounter when searching for free content. This is confirmed by industrial reports that highlight many of the cyberlockers we observe as key offenders~\cite{Behind2014}.
From these two data sources, we identified three regularly occurring websites: putlocker.is, watchseries.gs and vodly.cr~\cite{courtorder2013}. These three sites mainly index streaming links to movies, with an additional small fraction of TV shows. In this paper, we use the term $video$ to refer to both.
We emphasise that these may not be representative of \emph{all} indexing sites --- our analysis is specific to these three large sites, although we note these are significant players in the broader ecosystem.

%  We further solidified our choice by inspecing court orders obtained by the MPAA, which confirmed that both \ws{vodly.cr} and \ws{putlocker.is}) had been prosecuted, resulting in requests for ISPs to block them~\cite{courtorder2013}

% \gareth{Add context about what these website index. Movies, TV shows? When you refer to a "video" in the paper, what are you referring to? A movie?}

%\gianluca{can we back this up further?}

%These rank 330,235 (7,987 in AU), 159,301 (77,145 in US) and 31,995 (20,436 in US) respectively in Alexa. Note, however, that these ranks are deceptive as these websites tend to mirror themselves across many Top Level Domains (TLDs), thereby (technically) splitting their views across multiple sites listed in the Alexa rank.

We have designed a crawler that iterates over all video pages indexed on each of the three indexing sites. It extracts the video title, release year, genre and all associated streaming links. As previously stated, the indexing sites do \emph{not} host any content --- only links to external cyberlockers. We initiated this crawl on 12/01/2017 and repeated it on a monthly basis until 12/09/2017\footnote{\textit{putlocker.is} was crawled for monthly period starting 12/01/2017, 12/02/2017, 12/03/2017 as it went offline afterwards. In the case of \textit{vodly.cr}, we crawl it from 12/04/2017 onwards}. Table \ref{table:1} summarises the data for each indexing site targeted.

\subsection{Streaming Cyberlockers} 

After each monthly snapshot was gathered from the three indexing sites, the crawler followed each streaming URL to gain data from the cyberlockers themselves. We identified a total of \NumberOfCyberlockers~streaming cyberlockers on the indexing websites. We identify individual cyberlockers using their domain name; note that this includes mirrored cyberlockers with different Top Level Domains (TLDs). Unless stated otherwise, we treat these as different portals. The cyberlockers had diverse setups, and
many had taken steps that made crawling challenging. For example, six domains used Dean Edward's compression algorithm\footnote{\url{http://dean.edwards.name/weblog/2007/08/js-compression/}} for obfuscating the server hosting the content. As it was impossible to scrape all \NumberOfCyberlockers~cyberlockers,\footnote{Partly due to the frequency by which these websites change their web interface} we selected the \NumberOfStreamingDomain~most prominent streaming domains; this set covered \NumberOfStreamingLinkCrawalwdFurtherPercent~of extracted streaming links. The selected domains were those which were currently online and made the video information available to collect.
The domains that were not selected were either offline at the time of scraping or redirected to a different site. Unlike YouTube, we found that the user interfaces were quite primitive, lacking reliable metadata \eg view count and date of upload. For instance,  64\% of examined streaming cyberlockers did not allow searching and 42\% of portals ``curated'' their front-pages with legal short videos, which appear to have fake view counters. Therefore, for each video, we only recorded whether or not the video was online and the domain of the server it was hosted on. 
%Note that the video is always hosted on a different server to the HTML of the webpage. We focus on the video host, as this is where any illegal content is hosted.

\subsection{Cyberlocker Metadata}
Once we had collected all cyberlockers, we compiled metadata for each one. For every cyberlocker domain we performed DNS propagation checks around the world to generate domain $\rightarrow$ IP address mappings. We discovered a total of \NumberOfStreamingServers~distinct IP addresses hosting videos. We mapped each IP address into its geographical locations using Maxmind GeoLiteCity\footnote{\url{https://www.maxmind.com/}} and
Autonomous System (AS) using Team Cymru.\footnote{\url{http://whois.cymru.com}}
We discovered servers distributed across \NumberOfCountry~countries, \NumberOfContinent~continents and \NumberOfAutonomousSystem~distinct Autonomous Systems (ASes). Following this, we loaded all cyberlocker homepages using phantomjs.\footnote{\url{http://phantomjs.org/}} Upon each load, we recorded all the first and third party domains loaded by the page.

%This allowed us to identify trackers and ad networks used.

%For majority of the selected streaming domain, streaming server information is
% placed within Javascript, we employed three different methods to get this
% depending on which method that fits the domain. The first method we employed is
% using regular expression to extract desired information from video streaming
% page source HTML. However, some streaming domains deploy crawling prevention
% mechanism (Javascript obfuscation) to prevent any form of automated extraction.
% To get past this, we use same technique in the first method to get obfuscated
% text, we then deobfuscate it to extract streaming server information, this is
% the second method. Interestingly, all streaming domain we examined that employ
% obfuscation mechanism all use the same obfuscation algorithm - Dean Edward's
% compression algorithm \footnote{
% \url{http://dean.edwards.name/weblog/2007/08/js-compression/}}. In third method,
% we extract temporary stream link from streaming page using regular expression
% make a GET request to the temporary link and extract streaming server
% information from the location property in the response headers.

\subsection{Lumen Database}
A major theme in our work is understanding the role that video portals play in copyright infringement. It is, therefore, necessary to obtain ground truth data on which videos compromise copyright. To gather such data, we have scraped the Lumen database between 01/01/2017 and 30/09/2017 (the same period as our cyberlocker crawls). Lumen is a platform that aggregates legal complaints and requests for removal of online content. Each record covers an individual complaint to one or more organisations. An entry contains the URL(s), the complainant, the date and the complaint target (\ie{} a cyberlocker). Lumen predominantly captures complaints made to Google for removing content links from search results. Beyond this, Lumen also contains complaints to other search and social media sites, \eg Bing and Twitter.

% \begin{table}
%     \begin{tabular}{ll}
%     \hline
%     ~ & Jan - Sep \\ \hline
%     Number of take-down notices & \ref{label:total_lumen_notice} \\
%     Number of infringing URLs & \ref{label:total_infringing_url} \\ \hline
%     \end{tabular} \end{table}

%%%%%%%%%%%%%%%%%%%%%%%%%%%%%%%%%%%%%%%%%%%%%%%%%%%%%%%%%%%%%%%%%
%%%%%%%%%%%%%%%%%%%%%%%%%%%%%%%%%%%%%%%%%%%%%%%%%%%%%%%%%%%%%%%%%
\section{Characterising Indexing Portals} \label{sec:char_links}
%%%%%%%%%%%%%%%%%%%%%%%%%%%%%%%%%%%%%%%%%%%%%%%%%%%%%%%%%%%%%%%%%
%%%%%%%%%%%%%%%%%%%%%%%%%%%%%%%%%%%%%%%%%%%%%%%%%%%%%%%%%%%%%%%%%

When a user wishes to view a video, the first entity they must interact with is an \emph{indexing} site. In this section, we review what links are made available on these indexing portals, as well as the cyberlockers and content they point to.

%%%%%%%%%%%%%%%%%%%%%%%%%%%%%%%%%%%%%%%%
\subsection{How Many Links Are Available? }
%%%%%%%%%%%%%%%%%%%%%%%%%%%%%%%%%%%%%%%

\begin{figure}[t!] 
\includegraphics[width= 8cm, height=5cm]{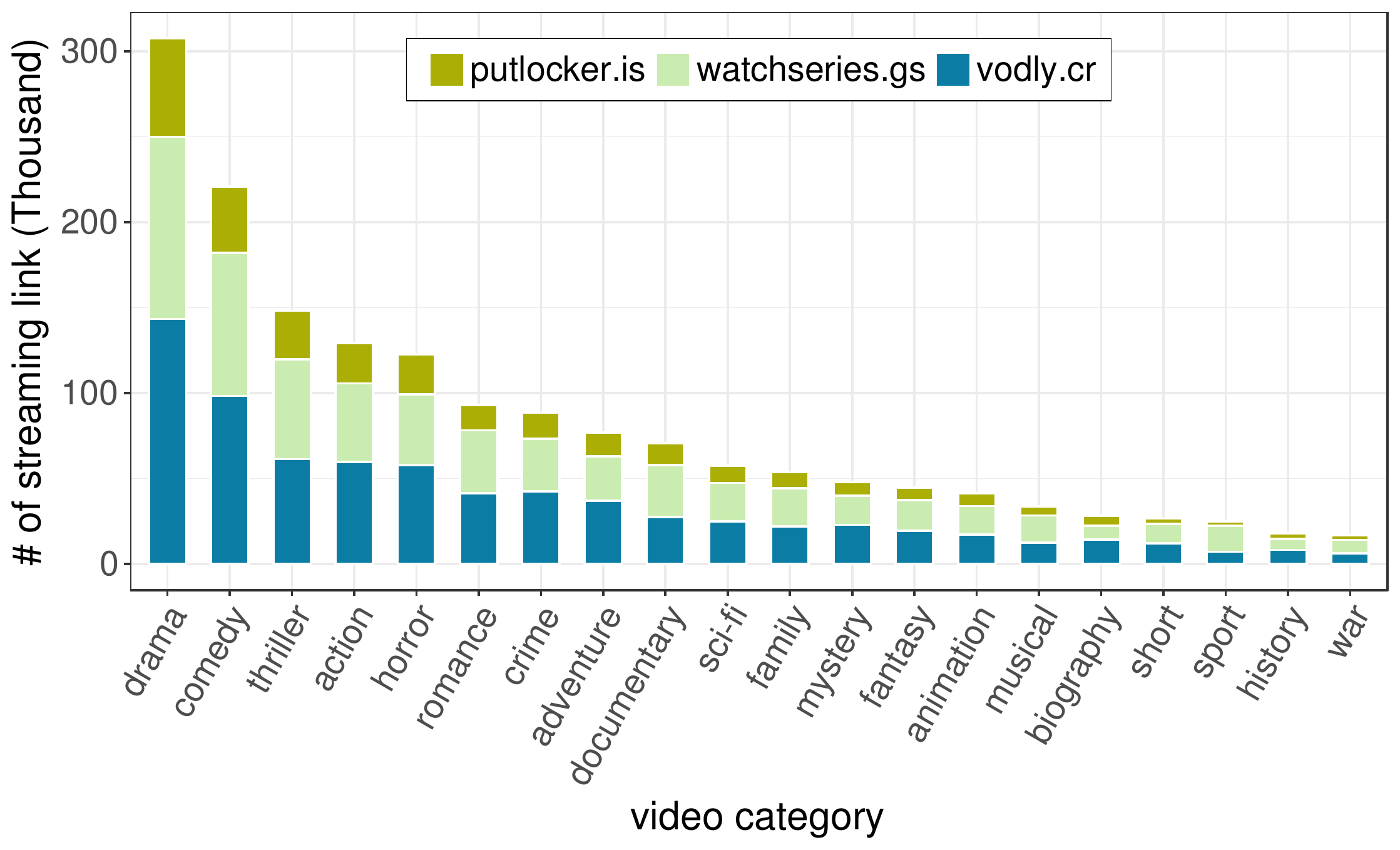}
\caption{Number of streaming links per category.}
\label{fig:genre}
\end{figure}

We begin by inspecting the \emph{number} of content items being indexed over time. This can be measured from two perspectives: \one~the number of video pages made available (there is one page per video) and \two~the number of streaming links made available on those pages. The former represents the number of new videos added to the indexing portals, whilst the latter captures the number of links per video. To give a brief understanding of the \emph{types} of videos available, Figure~\ref{fig:genre} shows the number of links within the top 20 genres specified on the indexing sites (this also coincides with IMDB's\footnote{http://www.imdb.com/} top 20 genres). It can be seen that Drama, Comedy, Thriller, Action and Horror videos dominate; the distribution in each indexing site is roughly equal and all follow an identical ranking.

\begin{figure}[t] \includegraphics[width= 8.5cm,
height=5cm]{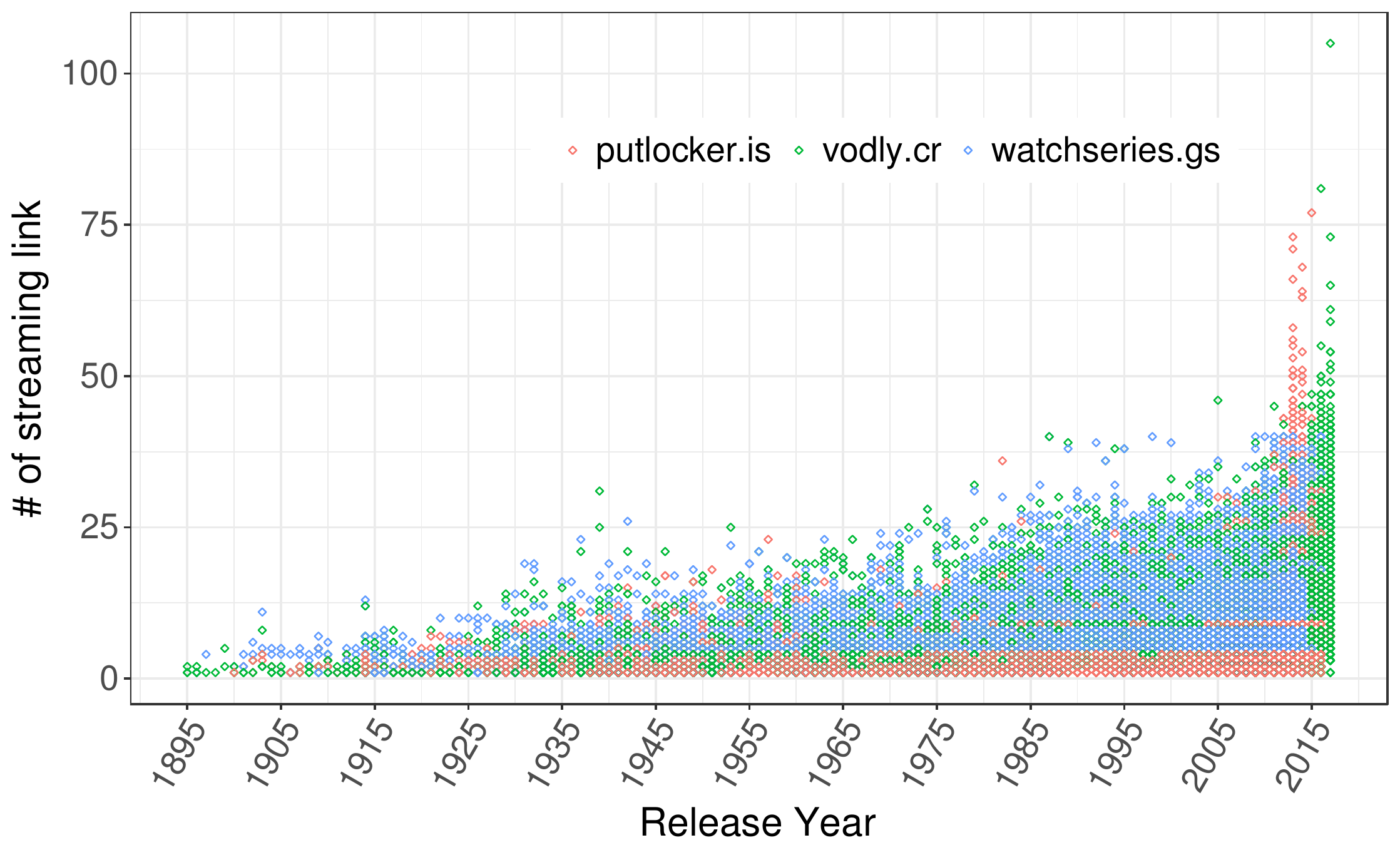} \caption{Number of streaming links
per video page. Video pages are split into release year.} \label{fig:indexing_site_streaming_links}
\end{figure}

When combining all genres we discover a total of \TotalNumberOfVideos\\video pages and \NumberOfStreamingLinksInDataset~streaming links. Figure~\ref{fig:indexing_site_streaming_links} plots the number of streaming links attached to each video for each indexing site (across each release year). On average a video has \AverageNumberOfStreamingLinkPerVideo~streaming links, but there is clear relationship between the recency of the release and the number of streaming links available. About 73\% of links are for videos released since 2000. Diversity can also be observed across the different portals: this figure is 81\% for \emph{putlocker.is}, 74\% for \emph{vodly.cr} and 69\% for \emph{watchseries.gs}. This indicates that the portals offer different styles of corpora. Overall, the average number of streaming links for videos with recent release years ($\geq$2000) is 7, compared to just 4 for earlier releases. 

We also observe that 7\% of video pages list \emph{no} streaming links; this suggests that either the links were removed, or the pages were generated without links being added. This is particularly prevalent for older videos.  About 11\% of videos with release years before 1980 do not have any streaming links, compared to just 6\% for later release years. Only 0.3\% of videos in 2017 have no links. This is likely driven by the higher demand and the more proactive participation of people uploading fresh content. That said, these portals also contain extremely old content, some over 100 years old.  Characterising these portals as exclusive copyright-infringement platforms may therefore be misplaced. 
%The oldest three videos are \textit{Tables Turned on the Gardener --- 1895}, \textit{The Arrival of a Train --- 1896} and \textit{A Terrible Night --- 1896}. 
Curiously, the fraction of films released before 1950 without streaming links is actually lower than later films  --- just 6\%. We assume this is because such videos are not aggressively pursued by copyright enforcers, hence reducing the number of legal actions.

%%%Removed: This is basically showing the same as the next figure - not worth the space unless conveying an important message % 
% \begin{figure}[t] \includegraphics[width= 8.5cm,
%   height=5cm]{indexing_site_streaming_link_availability} \caption{Breakdown of
%     percentage of videos with streaming links \vs those without based on the
%     video release year. Majority of videos have streaming links but those
%     without are usually of older release year, suggesting that recent videos are
%     guaranteed to have streaming links.}
% \label{fig:indexing_site_streaming_link_availability}
% \end{figure}

\subsection{Which Cyberlockers Are Most Popular?}
%%%%%%%%%%%%%%%%%%%%%%%%
 %Top 1M
 
\begin{figure}[t] \includegraphics[width= 8.5cm,
height=5cm]{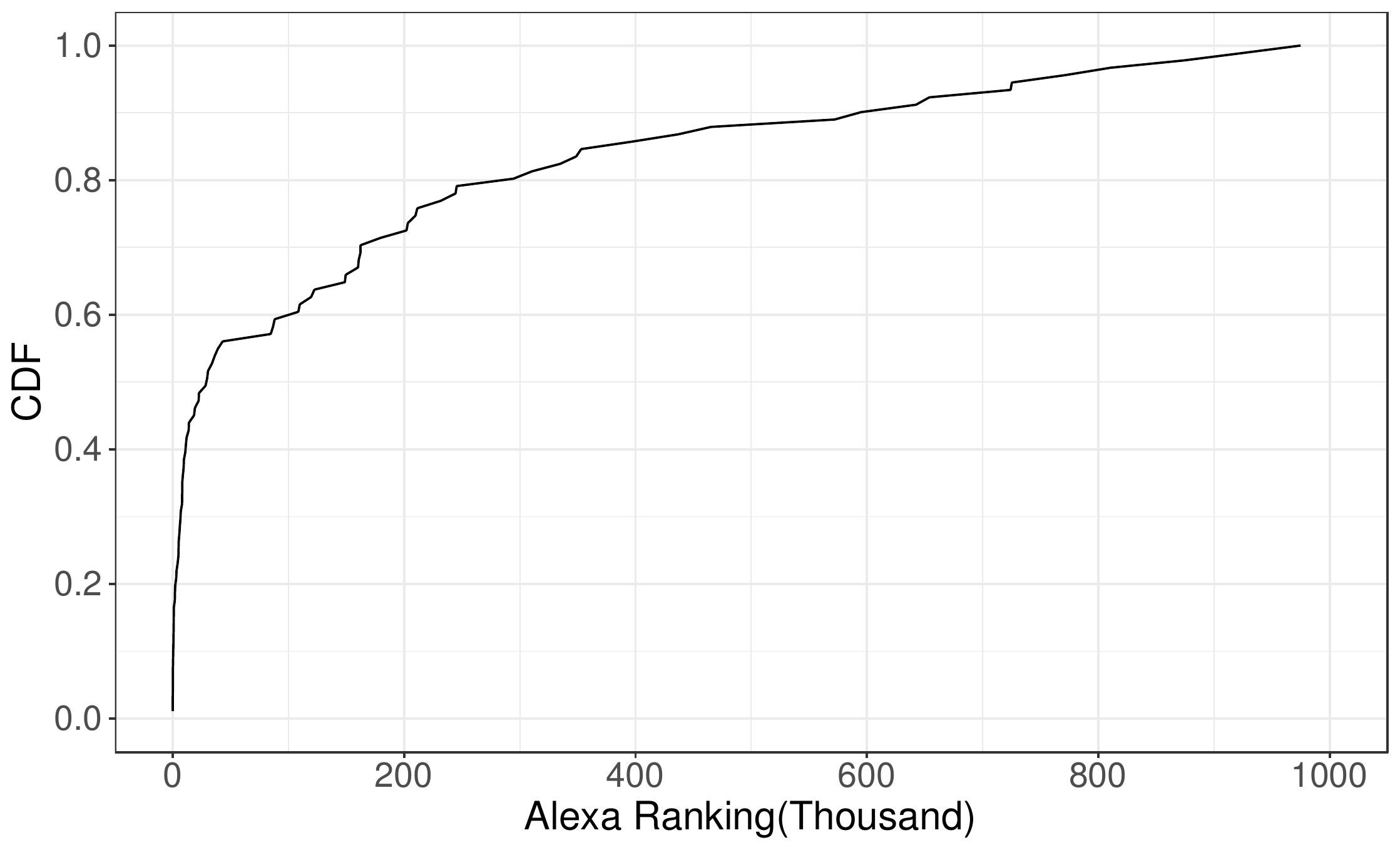} \caption{Cumulative Distribution Function (CDF) showing the distribution of streaming domains based on their Alexa Rank.} 
\label{fig:streaming_site_alexa_ranking}
\end{figure}

% \gareth{If we have time, it would be good to increase the size of all the graph fonts where possible. For example, the axis in the CDF and the key in Fig 2 are pretty small}
 
\begin{figure*}[ht] \includegraphics[width= 17cm, height=10cm]{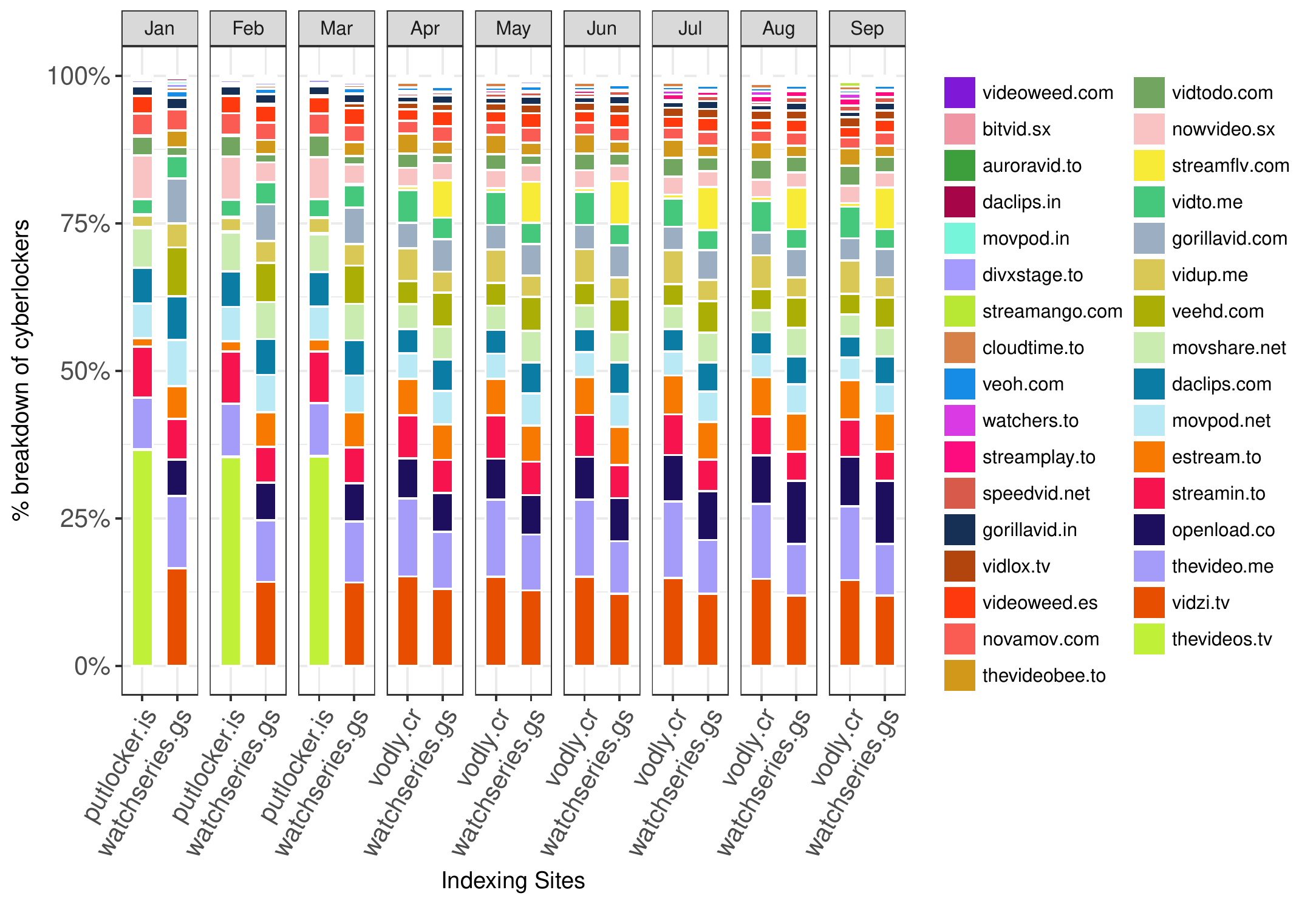}
\caption{Breakdown of streaming links seen on each indexing site per month. We began crawling indexing site \emph{vodly.cr} in April when \emph{putlocker.is} was taken offline. The stacked bar is ordered with the largest cyberlocker at the bottom.}
\label{fig:video_host_breakdown_timeseries} 
\end{figure*}

The previous section inspected the \emph{number} of streaming links. Next, we investigate \emph{which} cyberlockers are most prominent. From the \NumberOfStreamingLinksInDataset~streaming links extracted, there are \NumberOfCyberlockers~unique streaming cyberlocker domains. We first inspect their popularity as measured by the Alexa Rankings. Figure~\ref{fig:streaming_site_alexa_ranking} presents a Cumulative Distribution Function (CDF) of the Alexa ranks for the cyberlockers. About 60\% of these streaming domains are in Alexa's Top 1M. Amongst these, 70\% are in the Top 200K. The top three most popular streaming sites are \emph{openload.co} (rank 147), \emph{thevideo.me} (543) and \emph{vidzi.tv} (745). These rankings, however, do not correlate well with the number of videos hosted on the domain (Spearman coefficient of -0.015). For example, \emph{streamin.to} hosts 30,401 videos compared to just 7,288 for vidlox.tv and 1,924 for \emph{streamango.com}. Despite this, the latter two rank 5,699 and 2,124 compared to just 6,625 for \emph{streamin.to}. We can also inspect popularity through the lens of the indexing sites. Figure~\ref{fig:video_host_breakdown_timeseries} presents a breakdown of the streaming links that make up the indexing sites, split by monthly snapshot.  This is primarily intended to visualise the breakdown of cyberlockers per month, rather than their evolution over time. Note that the indexing sites vary across the time periods because \emph{putlocker.is} ceased uploading new content in April, to be replaced by \emph{vodly.cr}.

Firstly, it can be seen that well known user-generated content platforms such as YouTube, Vimeo or Dailymotion are not observed once. Instead, the indexing portals exclusively link to videos hosted on platforms that operate outside of the ``mainstream'', \eg \emph{thevideos.tv}, \emph{movpod.net} and \emph{videoweed.es}. Secondly, it can be seen that the cyberlockers present on each indexing site are different. This suggests communities where individual cyberlockers are associated with particular indexing sites. 30\% of cyberlockers are exclusive to a single index; 33\% are seen on two; the remainder appear on all indexing sites. The latter are, unsurprisingly, those with the greatest number of links. From the cyberlockers found on multiple indexing sites, 73\% of their links are unique and seen once. In other words, only 27\% of cyberlocker links are posted on more than one of our indexing sites. This suggests that different pirates have quite different strategies for promoting links to their content.

The prominence of each of these cyberlockers also changes across the monthly snapshots. For example, in February, we witness the introduction of \emph{vidlox.tv} and  \emph{streamplay.to}; in March --- \emph{streamflv.com}; in April --- \emph{speedvid.net}; in July --- \emph{watchers.to}  and in August --- \emph{streamango.com}. We also observe removals of cyberlockers, \eg in April, \emph{thevideos.tv} ceases to be indexed. This is because, prior to this, it was exclusively indexed by \emph{putlocker.is}. Upon ceasing operation in April, the loss of \emph{putlocker.is} meant that \emph{thevideos.tv} disappeared from our vantage point. 

We also see arrival and removal dynamics within individual links to each of the cyberlockers. Out of the~\NumberOfStreamingDomain~streaming cyberlockers we examined, we observed that 25 had links \emph{both} added and removed. The remaining 8 had only additional links injected, and never had any removed: these were \emph{openload.co}, \emph{vidtodo.com}, \emph{vidup.me}, \emph{estream.to}, \emph{streamplay.to}, \emph{vidlox.tv}, \emph{streamango.com}, \emph{watchers.to}. In total 55\% of cyberlockers saw growth during our measurement period, whilst 45\% saw a decline. The most extreme was \emph{divxstage.to} ,which in June had 24\% of its links removed from the indexing sites. In contrast, in July \emph{streamplay.to} saw a 107\% increase in the number of links indexed. These aggressive dynamics are presumably enabled by the ease that uploaders can move between cyberlockers. 

\section{Characterising Cyberlockers} \label{sec:relationships}
%%%%%%%%%%%%%%%%%%%%%%%%%%%%%%%%%%%%%%%%%%%%%%%%%%%%%%%%%%%%%%%%%%%%%%%%%%%%%%%%%%%%%
%%%%%%%%%%%%%%%%%%%%%%%%%%%%%%%%%%%%%%%%%%%%%%%%%%%%%%%%%%%%%%%%%%%%%%%%%%%%%%%%%%%%%
The above has revealed a wide range of cyberlockers. We next explore the characteristics of these cyberlockers, specifically regarding \one~the use of third party domains within the webpages; \two~the hosting of their video content; and \three~the similarities between the webpage HTML structure. Whereas the first two aspects shed light on the design and build of the websites, the third provides insight into potential relationships that may exist between the websites.

\begin{figure*}[tb] 
\includegraphics[trim=0cm 6cm 0cm 2cm, width= 18cm, height=6cm]{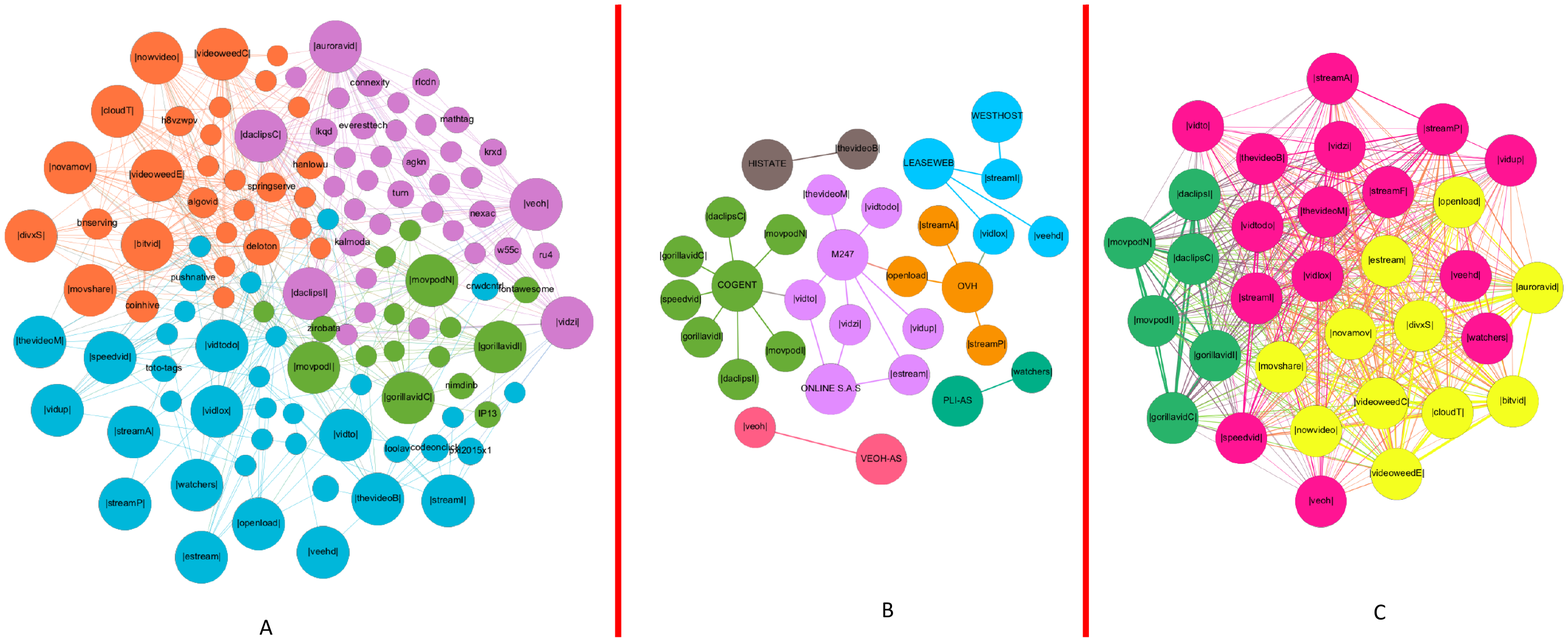}
\caption{$(a)$ Third-party domains linked by streaming cyberlocker homepages.\space$(b)$ Autonomous Systems where video servers are hosted.\space$(c)$ HTML similarity between cyberlocker homepages.
Within each sub-figure, each colour indicates a community formed when the $Louvain$ method for community detection is applied. Note - some hosting ASes used by some sites are not present within\space$(b)$ as these cyberlockers have not injected live videos during our crawl. In \space$(b)$ and\space$(c)$, the size of a node is proportional to its degree, while the thickness of an edge in\space$(c)$ indicates the closeness in similarity 
} 
\label{fig:streaming_site_relationship}
\end{figure*}

\subsection{Modelling Cyberlockers}

To model the relationships between cyberlockers, we embed them into a series of graph structures. Each graph captures shared characteristics and potential relationships between the websites. We use several techniques to generate three graphs from our data:

\begin{itemize}
 \item $Domains=(V, D, E_1)$, where $V$ is the set of cyberlockers, $D$ are third party domains, and $E_1$ link third party domains to the cyberlockers where they are embedded (in the homepage). As we are interested in identifying relationships, we filter domains with a degree of one. 
 
 \item $Networks=(V, N, E_2)$, where $N$ consists of Autonomous Systems (ASes),
and $E_2$ is a set of directed links indicating that a cyberlocker is hosted within a given AS. This allows us to reason over the hosting strategies of these operators.
 
 \item $HTML=(V, E_3)$, where $E_3$ contains weighted links based on the homepage HTML similarity between two cyberlockers $\in V$. Similarity is computed using the tag-based algorithm described in~\cite{cruz1998measuring}. For each pair of cyberlockers, we obtain a weight $t$ $\in$ $[0,100]$. A weight of 100 indicates identical HTML structures; a value of 0 indicates entirely disjoint HTML structures. The rationale for this is to reveal if there are some individuals who control multiple cyberlockers by simply reusing the same or similar website templates.
 
\end{itemize}

Once the graphs are generated, we use the $Louvain$ algorithm~\cite{Blondel2008} to perform graph clustering. This is intended to identify communities based on shared characteristics in $Domains$, $Networks$ and $HTML$. In the case of $HTML$, this explicitly highlights cyberlockers that were likely generated by the same operator.

\subsection{Understanding Cyberlockers}
%%%%%%%%%%%%%%%%%%%%%%%%%%%%%%%%%%%%%%%%%

By exploring these three graphs an understanding of individual cyberlocker characteristics can be gathered, as well as providing insite into why and where these features are shared. Figures~\ref{fig:streaming_site_relationship}(a)(b)(c) present the graphs of {\em Domains}, {\em Networks} and {\em HTML}; the nodes within these graph are coloured to indicate the individual communities detected using the Louvain algorithm.

%%%%%%%%%%%%%%%%%%%%%%%%%%%%%%%%%%%%%%%%
\subsubsection{\textbf{Shared Third-Party {\em Domains}}}
\label{sec:relationships:domains}
%%%%%%%%%%%%%%%%%%%%%%%%%%%%%%%%%%%%%%%%

Firstly, we look at the third party domains embedded within the cyberlockers. These include various domains ranging from ad networks to analytic services. In Figure~\ref{fig:streaming_site_relationship}(a) we identified 87 third-party domains (nodes) forming \textbf{4} communities. The globally most central nodes are three advertisement/tracking platforms --- google.com (betweenness of 1,580), rtmark.net (536) and deloton.com (376). Structurally these create a fully connected graph with most cyberlockers embedding these domains (or being a maximum distance of 2). Employing the Ghostery database\footnote{https://www.ghostery.com/} we were able to classify 60\% of extracted domains. Of these classified domains, 50\% was classified as \textit{First Party Exceptions}, 44\% as \textit{Advertising} and the rest split evenly between \textit{Patterns}, \textit{Site Analytics} and \textit{SocialMedia}. 

As the predominant form of monetisation, we next explore the domains classified as \textit{advertising}. Within these domains, popular ad networks include PopAds (in-degree 14), PubMatic (10) and DoubleClick (9). Note that PopAds specialises in ``popunder'' advertising~\cite{le2014attitudes} --- something banned by Google's AdSense. Generally though, these major ad networks forbid publishers with illegal content and, therefore, their terms of service are clearly being broken (which risks account removal). The large number (23) of advertisement brokers used by these cyberlockers suggests that these policies are not strictly enforced. If these ad domains were to cease offering adverts to the cyberlockers the operators would likely be significantly affected. With this in mind, we see several alternative monetisation tools beginning to emerge. More than a third of the cyberlockers have started using the recently released {\em Coinhive} plugin which mines cypto-currencies on the viewer's machine. Furthermore, we observe the presence of various malicious domains \eg \emph{codeonclick.com} (4 cyberlockers), \emph{rfihub.com} (3),  and \emph{nexac.com} (3), which download adware onto the viewer's machine, as well as \emph{mathtag.com} (2), \emph{exelator.com} (2) and \emph{btrll.com} (2), which perform browser hijacking. The dependency that cyberlockers have on these revenue sources suggests their removal would pose a severe risk to their operation, unlike most prior P2P platforms which are self-sustaining.

\subsubsection{\textbf{Co-located {\em Network} Hosting}}
\label{sec:relationships:networks}

Following on, we look at the Autonomous Systems (ASes) used to host video content. This is again modelled as a graph with links indicating the hosting of a particular cyberlocker within a given AS. This is important for several reasons. Most notably, the Digital Millennium Copyright Act (DMCA) allows legal parties to approach ASes who may have control over web content within their networks. These networks, therefore, represent a potential point of failure for the cyberlockers. In Figure~\ref{fig:streaming_site_relationship}(b), we see a total of \textbf{7} communities formed around such providers.

Firstly, we see three isolated communities: \emph{thevideobee.com} (hosted on HISTATE), \emph{veoh.com} (VEOH-AS) and watchers.to (PLI-AS). These jointly contribute only 3.4\% of the streaming links from our selected set of cyberlockers. Thus, the impact of removing these websites (or ASes) would be limited. The remaining four communities are larger and inter-connected via a series of bridge nodes. Bridge nodes are, by definition, those websites that spread their content across \emph{multiple} ASes. These are \emph{openload.co}, \emph{vidlox.tv}, \emph{streamin.to}, \emph{estream.to}, \emph{vidzi.to}, and \emph{vidto.me}).\footnote{Shutting down these streaming cyberlockers would remove 33\% of all videos.} Within the four inter-connected communities, M247 and Cogent have the highest degree centrality, with a degree of 7 and 8 respectively (\ie they host 7 and 8 cyberlockers each). Shutting down these two ASes would result in 58\% of the videos, and 71\% of the servers observed in our set becoming unavailable. This indicates a remarkable level of vulnerability and clearly a point of failure that could be leveraged by copyright enforcers. 
%Other notable players include ONLINE S.A.S, LEASEWEB and OVH, making make up 8\% of all available videos.

We posit that the 6 cyberlockers hosting content across multiple ASes may do so to increase resilience against takedowns (\S\ref{sec:removals:reactions}). Furthermore, to bolster this redundancy, there is a clear trend in the \emph{physical locations} of the servers. M247 is based in Romania, which (as a country) hosts the largest share of streaming servers, containing 42\% of the total streaming links witnessed. Similarly Cogent/Leaseweb are based in The Netherlands which hosts 23\% of streaming links. This trend is reportedly driven by the lax copyright enforcement within these countries combined with their high capacity Internet infrastructure~\cite{countries2013}. A sudden increase in copyright regulation within these countries may see a shift in this behaviour and, again, we argue that this dependency on individual countries poses a resilience challenge for the cyberlockers.

\subsubsection{\textbf{{\em HTML} structure of cyberlocker homepage}}
\label{sec:relationships:html}

Lastly, we compare the HTML of the cyberlocker homepages to detect underlying similarities between sites. This is because we hypothesise that some individuals may create \emph{multiple} cyberlocker front-ends (\eg to increase resilience). If this is true, it could imply that large segments of the cyberlocker system, which appear independent on the surface, are actually orchestrated by the same individual or organisation.  To compute this, we use an existing pattern matching algorithm~\cite{john1988}, which gives each pair of websites a similarity score, $t$ $\in$ $[0,100]$. To add context, we executed the algorithm on the Alexa Top 100 websites: the median similarity score was just 2.5. We then built a weighted graph with links between websites weighted by their similarity. 

The resulting weighted similarity graph is presented in Figure~\ref{fig:streaming_site_relationship}(c). The thickness of an edge indicates the similarity score.  However, unlike the previous bipartite graphs, the communities present show a \emph{direct} (rather than transitive) relationship between cyberlockers.  Within the graph two main communities can be identified, the Green group and the Yellow group. The median similarity scores in these two groups are 77.9 and 53.6 respectively. The Pink group contains the remaining very loosely connected cyberlockers with a median similarity of just 18.3. Manual inspection suggests that any scores above 30 indicate strong similarity.

It should also be noted that these similarities are also mirrored across $Domains$ and $Networks$. For example, the Green group all host within the Cogent AS.  We further observe that 4 out of 6 sites (\emph{gorillavid.com}, \emph{gorillavid.in}, \emph{movpod.net} and \emph{movpod.in}) fall into the same community within $Domains$. These inferences are confirmed by WHOIS,\footnote{A protocol for querying the registered owner of an Internet resource.} which reveals that \emph{gorillavid}, \emph{movpod}, \emph{daclips} are all registered with the same owner. Unfortunately, 48\% of our cyberlockers use WHOIS anonymisers to avoid registering their details. This prevents us from definitively proving shared ownership in other case. However, we briefly note that usage of anonymisers may itself indicate a similar owner. For instance, 0\% of the sites in the Green group utilise anonymisers, in contrast to 73\% in the Pink group. 

%If the gorillavid community were removed \tofillp of content would be deleted, whereas this would be \tofillp for the videoweed community. In total, if these two entities were taken down, \tofillp of servers and \tofillp of content would be removed. 

It is also important to understand why these similarities emerge. In some cases, the similarities are driven by websites using similar templates. For example, \emph{daclips}, \emph{gorillavid} and \emph{movpod} use the same structural template, but with different colour coding and other minor differences. The opposite extreme also exists --- \emph{novamov.com} operates as an front for \emph{auroravid.to}; all visits to \emph{novamov.com} redirect to \emph{auroravid.to}. From the 28 node pairs identified as having a similarity scores above 75\%, we find that 8 are identical pages and 8 have shared templates; the remaining 12 node pairs could be considered to share templates, but one or both of the nodes redirects to another page. From the identical pages, 3 are mirrored TLDs (\eg .com and .in), whereas the remainder have totally different domains but pointing into a shared page. Overall, these observations mean that at least one fifth of our cyberlocker domains are actually operated by just two organisations/individuals, again confirming a remarkable dependency on just a small number of stakeholders.

% \gareth{Why isn't videoweed group in the AS map - did all of these sites not having videos to check the AS of? \dami{They are in the same family as auroravid, dixstage and likes}}

%Finally, we briefly note that the make-up of these home pages differs quite substantially from mainstream video portals. 

% \begin{figure}[t]\includegraphics[width= 8cm, height=8cm]{cor/cor-ss-circle}
% \caption{Pairwise similarity among home pages (HMTL) of cyberlockers.}
% \label{fig:cor-ss-circle}
% \end{figure}

%%%%%%%%%%%%%%%%%%%%%%%%%%%%%%%%%%%%%%%%%%%%%%%%%%%%%%%%%%%%%%%%%%%%%%%%%%%%%%%%%%%%%
%%%%%%%%%%%%%%%%%%%%%%%%%%%%%%%%%%%%%%%%%%%%%%%%%%%%%%%%%%%%%%%%%%%%%%%%%%%%%%%%%%%%%
\section{Exploring Take down Dynamics} \label{sec:removals}
%%%%%%%%%%%%%%%%%%%%%%%%%%%%%%%%%%%%%%%%%%%%%%%%%%%%%%%%%%%%%%%%%%%%%%%%%%%%%%%%%%%%%

Finally, we briefly turn to our Lumen dataset to understand the level of copyright infringement taking place and the reactions of the websites to take down notices. 

%%%%%%%%%%%%%%%%%%%%%%%%%%%%%%%%%%%
\subsection{Overview of Complaints} \label{sec:removals:overview}
%%%%%%%%%%%%%%%%%%%%%%%%%%%%%%%%%%%
In total we find 780M infringing URLs from our crawl of the Lumen database.
This covers copyright complaints lodged from 01/01/17 to 30/09/17 and lists websites ranging from cyberlockers to Torrent sites. This list was, therefore, filtered to only include the \NumberOfStreamingDomain~streaming domains observed in our crawl. This left 21.8M infringing URLs across \NumberOfNoticeFound~individual complaints from \NumberReportingOrganisationFound~organisations.

\begin{figure}[tb!] \includegraphics[width= 8.5cm, height=5cm]{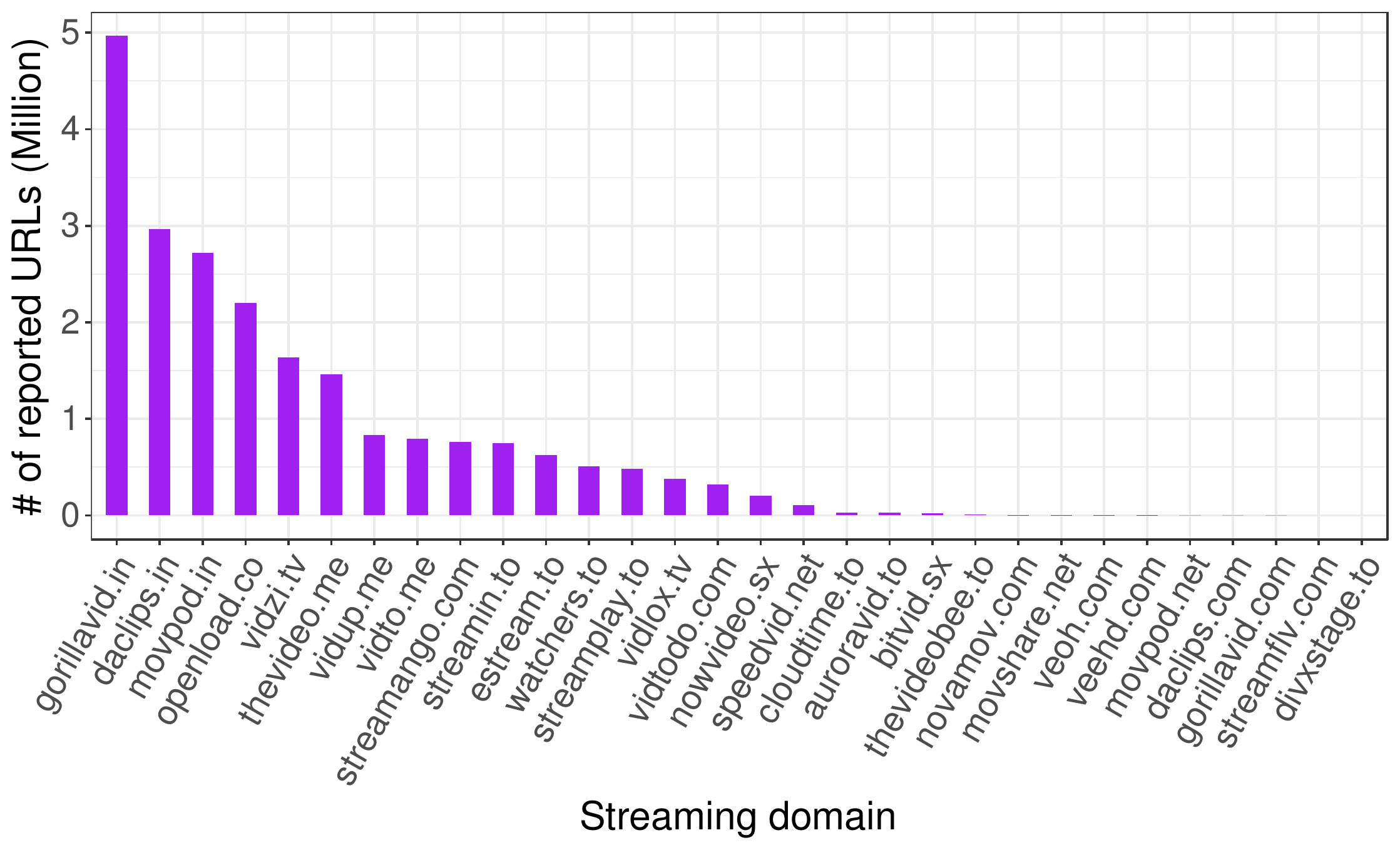}
\caption{Number of video streaming URLs submitted as infringing belonging to
streaming domains (Jan-Sep 2017). 
} \label{fig:reported_urls} \end{figure}

%\gianluca{any idea on why this is the case?}

Figure~\ref{fig:reported_urls} presents the number of complaints against each of the cyberlockers under-study. It can be seen that \emph{gorillavid.in} has the most complaints by far, followed by \emph{daclips.in} and \emph{movpod.in} (note, these three were identified as existing in the same community in both the $Networks$ and $HTML$ graphs). These sites account for 48\% of the total complaints made against our selected cyberlockers. Despite this, they do not constitute a major contributor to videos within our dataset (just 1.2\%, 0.3\% and 0.3\% respectively). We see more contributions from their mirrored websites under different Top Level Domains (TLDs): \emph{movpod.net}, \emph{daclips.com} and \emph{gorillavid.com} (4.7\%, 4.5\% and 3.8\%). These mirrored sites received far fewer complaints, despite possessing more links. When combined, the top 3 most popular sites (\emph{openload.co}, \emph{thevideo.me} and \emph{vidzi.tv}) receive 24\% of all complaints. %We find little correlation between the number of videos on a site and the number of complaints against it (Spearman -0.015), but the correlation with the Alexa ranking is much stronger (-0.81).

\begin{figure}[t!] 
\includegraphics[width= 8.5cm, height=5cm]{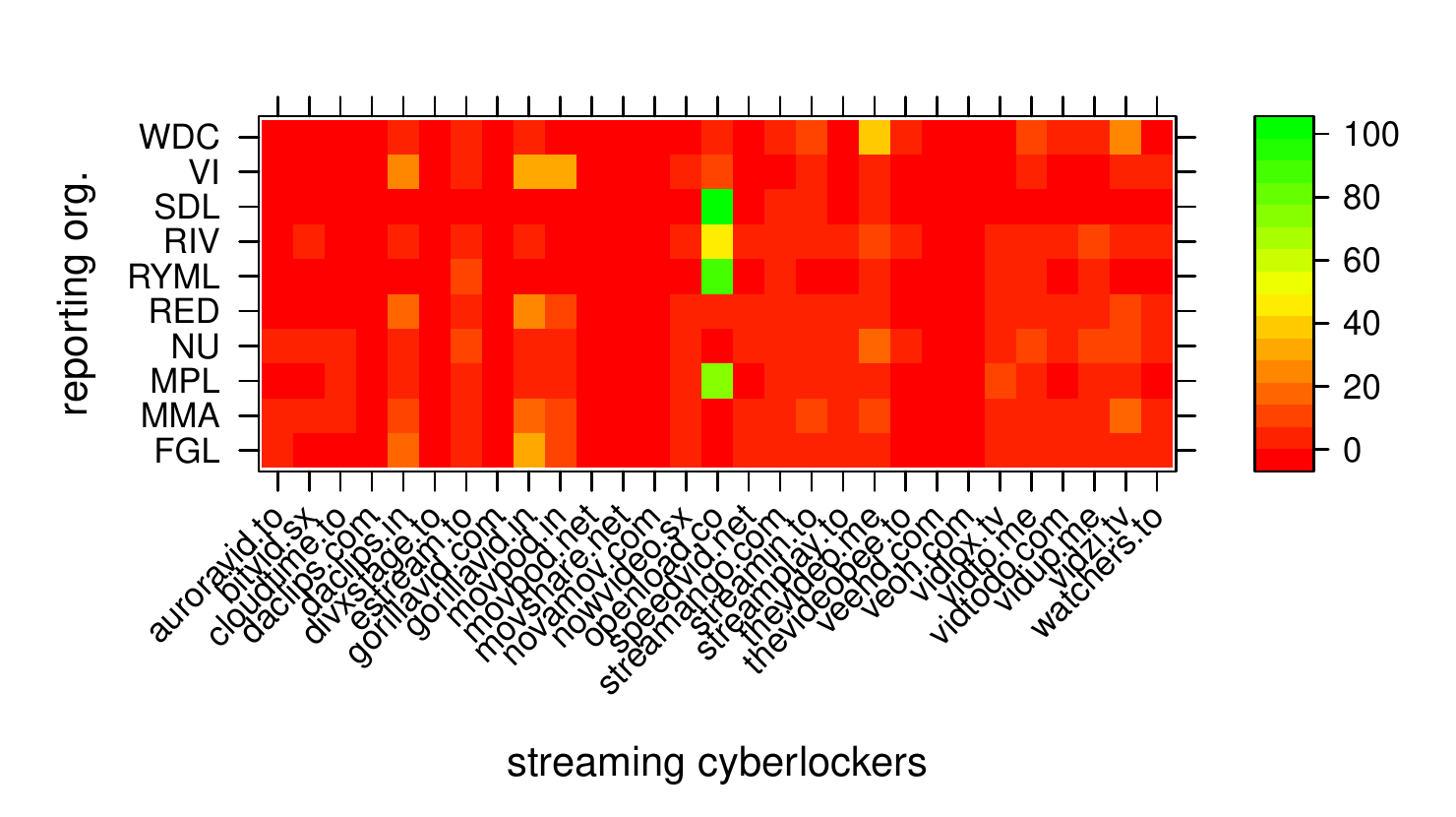}
\caption{Percentage distribution of infringing URLs lodged by reporting organisations against each cyberlocker} 
\label{fig:heat_map_copyright} 
\end{figure}

This leads onto the question of who generates these complaints? We identify a total of 304 notice senders --- Table~\ref{table:notice_sender} shows the top 10. Unsurprisingly, these are dominated by content producers such as 21st Century Fox and Walt Disney. We also find a number of dedicated anti-piracy companies (\eg Mark Monitor, Rivendell, Vobile). These top 10 notice senders contribute 96\% of all URLs complained about within the list examined, with the remaining 294 covering just 4\%. Upon closer inspection, trends can be observed among these top complainants. Figure~\ref{fig:heat_map_copyright} presents a heat map; the Y-axis list the top 10 complainants, the X-axis lists the cyberlockers. The heat represents the fraction of notices from each complainant to each cyberlocker. It can be seen that most complainants are highly selective in terms of which cyberlockers they complain about. For example, about 98\%, 89\% and 73\% of URLs complained about by \textit{Skywalker Digital}, \textit{Remove Your Media} and \textit{MG Premium Limited} were aimed at \emph{openload.co}. Why such organisations choose to target individual cyberlockers in this way is unclear. However, the trend generalises across other complainants too. For instance, we see \emph{daclips.in}, \emph{gorillavid.in} and \emph{movpod.in} being jointly targeted by \textit{Fox Group} (63\%), \textit{Mark Monitor} (38\%), \textit{Redacted} (55\%) and \textit{Vobile} (86\%). Despite this, we find no evidence that these cyberlockers contain more or less content belonging to each complainant. 

% 62.93362861	fox
% 37.99618617	MM
% 55.31775115	RE
% 86.24072317	VO

\begin{table}[t]
    \begin{tabular}{llll}
    \hline
    \begin{tabular}[c]{@{}l@{}}Notice\\ sender\end{tabular} &
\begin{tabular}[c]{@{}l@{}}No. of \\ URL\end{tabular} &
\begin{tabular}[c]{@{}l@{}}No. of \\ notice\end{tabular} &
\begin{tabular}[c]{@{}l@{}}No. of \\ domains\end{tabular} \\ \hline
     fox group legal & 8,508,289 & 1,876 & 22 \\
     markmonitor antipiracy & 5,777,393 & 5,795 & 24 \\
     rivendell & 3,369,745 & 9,842 & 26 \\
     vobile inc & 3,214,057 & 12,044 & 28 \\
     redacted & 400,329 & 228 & 28 \\
     remove your media llc & 291,213 & 488 & 21 \\
     nbcuniversal & 236,969 & 345 & 29 \\
     walt disney company & 168,139 & 1,472 & 25 \\
     mg premium ltd. & 97,176 & 1,063 & 26 \\
     skywalker digital ltd. & 89,046 & 649 & 22 \\ \hline
    \end{tabular}
     \caption{Top 10 copyright infringing notice senders}
    \label{table:notice_sender} 
\end{table}

\subsection{How Do Cyberlockers React?} \label{sec:removals:reactions}
%%%%%%%%%%%%%%%%%%%%%%%%%%%%%%%%%%%%

The above shows that complaints are regularly made against these portals. Next we inspect the reaction of cyberlockers to such complaints. Note, Lumen does not record specific complaints made \emph{to} the cyberlockers, they record complaints made \emph{about} them (to other parties \eg Google, Bing). Utilising the monthly snapshots, we can see if videos uploaded in 2017 were removed after a complaint had been lodged. Thus, we extract the set of complaints that correspond to videos in our dataset. This leaves a total of \NumberOfStreamingLinkIdentifiedInLUmen~streaming links reported in 2017, associated with \NumberOfVideosIdentified~videos released during this period. This, of course, leaves a large number of complaints in our Lumen database that we do not have corresponding video data for. This, unfortunately, is inevitable due to the sheer scale of the ecosystem.

\begin{figure}[t!] \includegraphics[width= 8.5cm,
height=5cm]{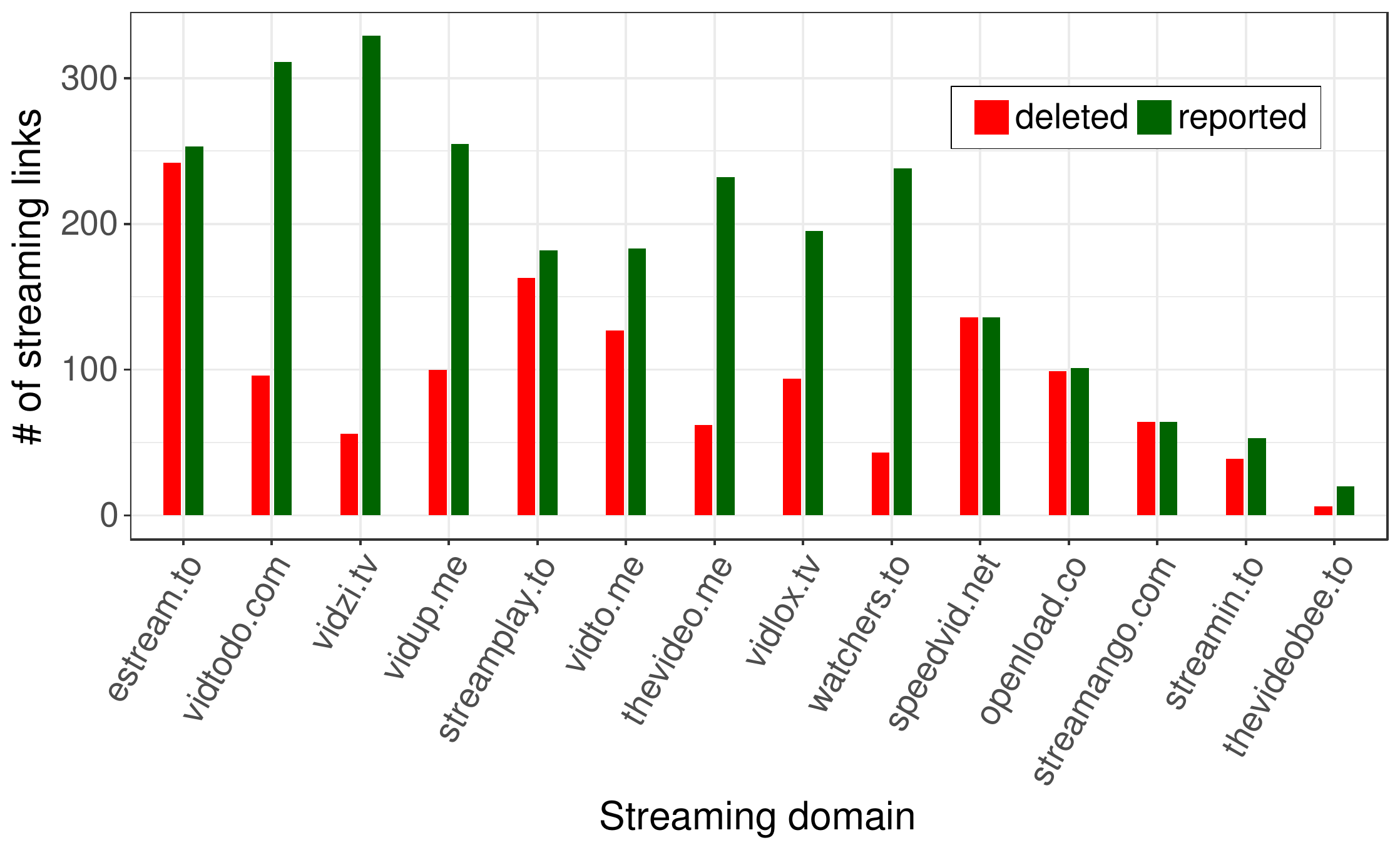} 
\caption{Cyberlocker URLs reported for infringing copyright compared to URLs deleted. X-axis ranked by fraction of takedown requests acted upon.} 
\label{fig:reported_available_deleted}
\end{figure}

% \gareth{From Gianluca: we should explain why we only are listing these cyberlockers, and not the others\dami{It has been added. 2nd sentence in 2nd paragraph section 6.2 }}

Figure~\ref{fig:reported_available_deleted} plots the number of complaints and the number of removals for each cyberlocker across our entire measurement period. Only cyberlockers which received requests to remove links gathered during our monthly crawls are included in the figure. Within the figure, if all videos complained about were removed, then the number of removals and the number of complaints would be equal.  The X-axis is ordered by the fraction of complaints acted upon. A clear ranking can be seen with some cyberlockers removing nearly all videos complained about\footnote{Note we cannot conclude that the removal was directly caused by the takedown notice. We can only assume this is likely.} (\eg \emph{estream.to}), whereas others (\eg \emph{vidzi.tv}) ignore nearly all complaints. We can compare this ``obedience'' rank against the others previously discussed. We find little correlation between this and the Alexa rank (Spearman Rank -0.015), but a stronger correlation with the number of links on the site (-0.81). This might exists because larger sites find it more difficult to ignore legal pursuit.

\begin{figure}[t!] 
\includegraphics[width= 8.5cm, height=6.3cm]{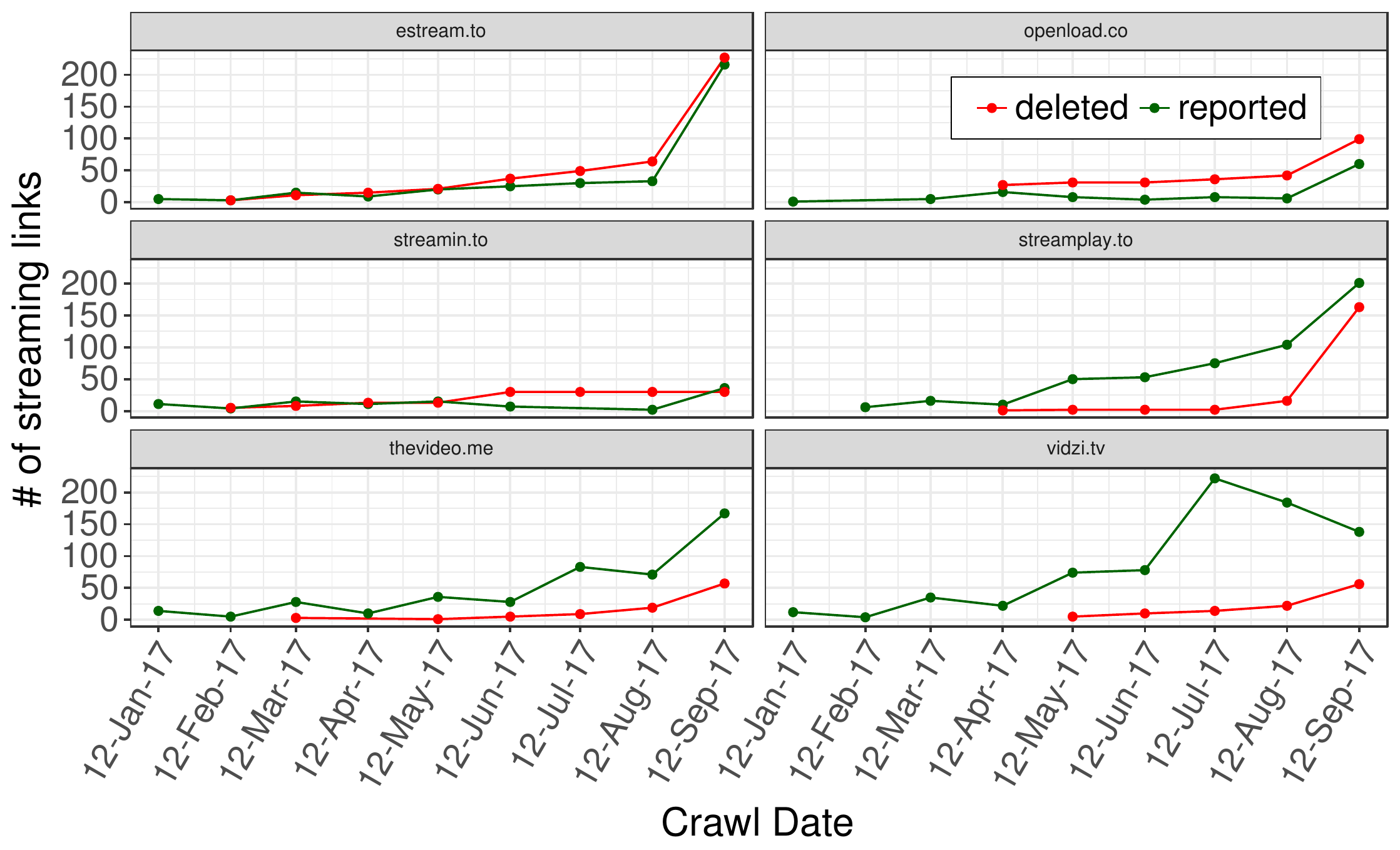} 
\caption{Time-series of removals. A reported link is available if at the month of crawl we can access the video.}
\label{fig:timeseries_reported} 
\end{figure}

To expand on this we can also explore the removals over time. Due to space constraints we select 6 cyberlockers that have a mixture of behaviours. For these we plot the number of videos reported and deleted over time in Figure~\ref{fig:timeseries_reported}. We observe a variety of behaviours. For example, websites such as \emph{openload.co}, \emph{estream.to} and \emph{streamin.to} react positively to copyright reports: over 75\% of videos are deleted within 1 month of complaints being registered on Lumen. The same cannot be said of \emph{vidzi.tv} and \emph{thevideo.me}, where $<$30\% of videos are deleted within 1 month. We observe that the videos that are \emph{not} deleted from \emph{openload.co}, \emph{estream.to}, \emph{vidzi.tv} are all hosted in Romania on M247 (the videos that are deleted are in other ASes).
%In total, 69\% of videos complained about within Romania get deleted. 
That said, it would be unwise to draw conclusions here, as Romania hosts both the cyberlocker that ignores the most complaints \emph{and} the cyberlocker that acts upon most complaints. Overall, the country hosting content that least frequently respects complaints is the Netherlands, where only 6\% of requests are acted upon. Hence, the diversity seen within individual countries suggests that the decision to act upon a complaint is largely driven by the individual cyberlockers.

\section{Conclusion \& Future Work} \label{sec:conclusion}
In this paper, we have offered a first study of the emerging streaming cyberlocker ecosystem. We began by exploring the streaming links shared on indexing sites (\S\ref{sec:char_links}). We discovered an environment actively involved in copyright infringement with an aggressive injection of recent releases. We proceeded to examine the individual characteristics and potential relationships between these websites (\S\ref{sec:relationships}). This identified a number of  communities based on shared domains, shared hosting facilities and high levels of HTML similarity. In some cases we found that this was individual operators running multiple cyberlocker instance or simply redirecting into the same (or mirrored) websites. This may be done for many reasons, but we believe it is most likely to increase resilience in the face of legal action (\S\ref{sec:removals}). Indeed, a common observation is the vulnerability of the cyberlockers. For instance, we find that over half of all content observed is hosted within just two ASes. 

This is just the first step in our exploration of the cyberlocker ecosystem. There are a number of future lines of work we will explore.
We emphasise that our data has only inspected a \emph{slice} of the streaming cyberlocker ecosystem. There are many other indexing sites, as well as portals that we have not investigated yet. This is evidenced by the number of complaints on Lumen that we did not have the corresponding video data for. Hence, our major line of future work is expanding our datasets to generalise findings across a broader swathe of the ecosystem.
We also wish to further investigate the \emph{relationships} between cyberlockers. This will involve building more graph-based models, underpinned by alternative data such as feature extraction from HTML. We further plan to investigate longitudinal trends. For example, we believe websites may periodically ``rebrand'' after they have previously been taken down --- studying this as an evolving series of graphs will help identify this. We also wish to better understand the evasion tactics used by these sites, particularly in the face of changing strategies used by copyright enforcers. This will, of course, involve deep diving into the way that enforcers select websites to complain about. %Lastly, we wish to investigate the popularity of videos hosted on cyberlockers. This is to determine if the popularity of a video influences the Alexa rank of the streaming cyberlocker it is uploaded on. 

% It is always good to cite \cite{li2009pirate} and acknowledge

\bibliographystyle{aaai}
\bibliography{mybibliography}

\end{document}